\documentclass[pra,twocolumn,amsfonts,floatfix,superscriptaddress]{revtex4}

\usepackage[dvips]{graphicx}  
\usepackage[latin1]{inputenc} 
\usepackage{amssymb}          
\usepackage{amsmath}          
\usepackage{amsfonts}
\usepackage[usenames,dvipsnames]{xcolor}
\usepackage{curves}
\usepackage{comment}

\newtheorem{lemma}{Lemma}

\newtheorem{defin}{Definition}
\newtheorem{thm}{Theorem}
\newtheorem{cor}{Corollary}

\newcommand{\proof}{\noindent {\em Proof. }}

\newcommand{\ket}[1]{|#1\rangle}
\newcommand{\bra}[1]{\langle #1|}

\newcommand{\Hi}{\mathcal{H}}

\newcommand{\Ei}{\mathcal{E}}
\newcommand{\trace}{\mathrm{{Tr}}}
\newcommand{\supp}{\textrm{supp}}

\newcommand{\compl}[1]{\overline{#1}}

\newcommand{\Prob}{\mathbb{P}}

\newcommand{\Di}{\mathfrak{D}}
\newcommand{\Ni}{\mathcal{N}}

\renewcommand{\span}{\textrm{span}}

\newcommand{\qed}{\hfill $\Box$ \vskip 2ex}

\renewcommand{\cal}{\mathcal}

 \newcommand{\beq}{\begin{equation}}
 \newcommand{\eeq}{\end{equation}}
 \newcommand{\beqa}{\begin{eqnarray}}
 \newcommand{\eeqa}{\end{eqnarray}}
 \newcommand{\beqan}{\begin{eqnarray*}}
 \newcommand{\eeqan}{\end{eqnarray*}}
 \newcommand{\bea}{\begin{eqnarray}}
 \newcommand{\eea}{\end{eqnarray}}

\date{\today}

\begin{document}


\title{Symmetrizing Quantum Dynamics\\ Beyond Gossip-type Algorithms}

\author{Francesco Ticozzi}                                               
\affiliation{Dipartimento di Ingegneria dell'Informazione, Universit\`a di Padova, via Gradenigo 6/B, 35131 Padova, Italy} \affiliation{Department of Physics and Astronomy, Dartmouth College, 6127 Wilder, Hanover, NH (USA). }\email{ticozzi@dei.unipd.it}  %

\begin{abstract} 
Recently, consensus-type problems have been formulated in the quantum domain. Obtaining average quantum consensus consists in the dynamical symmetrization of a multipartite quantum system while preserving the expectation of a given global observable. In this paper, two improved ways of obtaining consensus via dissipative engineering are introduced, which employ on quasi local  preparation of mixtures of symmetric pure states, and show better performance in  terms of purity dynamics with respect to existing algorithms. In addition, the first method can be used in combination with simple control resources in order to engineer pure Dicke states, while the second method guarantees a stronger type of consensus, namely single-measurement consensus. This implies that outcomes of local measurements on different subsystems are perfectly correlated when consensus is achieved. Both dynamics can be randomized and are suitable for feedback implementation.%
\end{abstract}
\maketitle

\section{Introduction}
In quantum as in classical systems, symmetry is a fundamental concept, and a key tool in investigating dynamical systems.
In particular, suitable dynamical symmetries, or equivalently the existence of preserved quantities, prevent controllability for quantum dynamics \cite{altafini-tutorial,dalessandro-book,thomas-controllability}, while they allow for the existence of protected sets of states \cite{knill-qec,viola-generalnoise,zanardi-symmetrizing,viola-dd}.
Symmetric quantum states and subspaces have a key role in the description of quantum systems obeying Bose-Einstein statistics \cite{kardar}, they are related to thermalization \cite{eisert}, and have a key role in quantum information \cite{nielsen-chuang}. Prototypical states used to illustrate and exploit intrisecally quantum correlations between information units, or {\em entanglement between qubits} in the quantum information jargon, are the maximally entangled states named after Greenberg-Horn-Zeilinger (GHZ) \cite{GHZ} and the W states \cite{W}: both are invariant with respect to permutation of their subsystems.

In \cite{mazzarella-group} it is shown that a class of classical dynamics obtaining asymptotic consensus can be seen as symmetrizing dynamics, leading to permutation-invariant states. The same underlying idea led to dynamics for symmetrizations in the quantum realm \cite{mazzarella-qconsensus}. There, it is shown how quantum consensus algorithms can be used in combination with simple local controls and measurements in order to prepare pure states and estimate the size of a network. This type of symmetrizing dynamics and their convergence properties, as well as their continuous-time counterpart, have been further studied in \cite{mazzarella-MTNS,petersen-qconsensus,petersen-qconsensus2}. All the symmetrizing dynamics that have been proposed so far, however, {are based on combinations of permutation operators, and hence they share two common properties: (1) they are {\em unital}, i.e. the maximally mixed state is preserved; (2) they attain symmetric state consensus, which is effectively consensus on the {\em statistical} properties of the variables of interest, but there is no algorithm that can attain actual consensus on the output of each local measurement. Due to the contraction properties of the considered maps, unital dynamics entails that the purity of the quantum state can not be augmented by the consensus-reaching dynamics.} With a degradation of purity, the fragile quantum correlations encoded in the state are typically lost, and stronger notions of consensus are out of reach. If the consensus-achieving dynamics improve purity, they can be instrumental to distributed and robust preparation of interesting states, like Dicke (including W) or GHZ states \cite{ticozzi-QL,ticozzi-QLS}, {as we shall demosntrate in the following.}

In the effort of overcoming these issues, two types of symmetrizing dynamics are proposed, which allow for asymptotic symmetrization of the state of a multipartite quantum system with respect to the permutation group acting on the subsystems. Both dynamics are composed of quasi-local maps and are robust with respect to randomization, and are hence suitable for distributed and unsupervised implementations -- as are their classical and classically-inspired counterparts. 

The first one attains an asymptotically symmetric state while preserving the expectation value of a global observable, reaching {\em average symmetric state consensus} in the language of \cite{mazzarella-qconsensus}. It does so by selecting and preparing a pure symmetric state (a {\em Dicke} state \cite{dicke,dicke-rev}) in each eigen-subspace of the observable of interest. { It is also shown how these dynamics are instrumental to preparation of globally entangled pure states. The proposed protocol uses only single system operations and pairwise interactions.}

The second dynamics is the first proposal of a quasi-local protocol that obtains a stronger type of quantum consensus, called single-measurement consensus \cite{mazzarella-qconsensus}. This type of consensus, which implies the symmetry of the state but for which the latter is not sufficient, is the closest in spirit to classical consensus: after single measurement consensus is reached, the measurement of a local observable quantity on any subsystem will force the whole network of systems to ``agree'' on the result, i.e. yielding perfectly correlating results. Notice that, in contrast with classical consensus, the consensus value may not be determined before it is actually measured. 

Both methods rely on essentially quantum features of the system and, while obtaining a symmetric pure state that has a specified average of an observable is not viable in general, they allow for a final purity that is typically better than the one offered by gossip-type algorithms. Both are  suitable for implementation via discrete-time feedback \cite{bolognani-arxiv}, and can be combined with local initialization procedures in order to actually {\em prepare perfectly pure and entangled symmetric states}. This is explicitly shown for the first proposed algorithm (see Corollary \ref{cor:prep}). These dynamics can be seen as the discrete-time equivalent of conditional preparation of entangled states, in the spirit of \cite{ticozzi-QLS}. 

The paper structure is as follows: in Section \ref{sec:background}  a brief review of the relevant quantum consensus definitions and the ideas underlying existing consensus-achieving dynamics is provided; Section \ref{sec:results} begins by explaining why trying to obtain quantum consensus with just pure states is impossible, and continues by providing the form and the convergence proofs for two novel symmetrizing dynamics.   While these can be straightforwardly extended to networks of $d$-dimensional systems, as those considered in the introductory Section, the presentation here focuses on qubit networks and pairwise interactions. Qubit systems are easier to visualize and yet relevant for applications, some of which have been discussed in \cite{mazzarella-qconsensus}. Pairwise interactions are the minimal that can be allowed for interacting dynamics, and if a more forgiving locality constraint is in place, a set of effective pairwise interactions will also be allowed under this locality notion. Section \ref{sec:example} illustrates the behavior of the two dynamics, and compares them with the existing consensus algorithm.

\section{Background} \label{sec:background}
\subsection{Quantum Consensus States}\label{sec:consensus}

Consider a multipartite system composed of $m$ isomorphic subsystems, labeled with indices $i=1,\ldots,m,$ with associated Hilbert space  $\Hi^m:=\Hi_1\otimes \dots \otimes \Hi_m\simeq \Hi^{\otimes m}$, with $\dim(\Hi_i)=\dim(\Hi)=n$ and $2\leqslant n < \infty$. This multipartite system will act as our {\em quantum network}. We shall use Dirac's notation: $|\psi\rangle$ denote vectors of $\Hi,$ $\langle \psi|$ denote their dual linear functional. $\mathfrak{B}(\mathcal{H})$ denotes the set of linear operators on $\Hi,$ which in our finite-dimensional setting are in a one-to-one relationship with complex matrices. States are associated to density operators, namely linear, trace-one, positive-semidefinite operators on $\Hi,$ with their set denoted by $\mathfrak{D}(\Hi)\subset\mathfrak{B}(\mathcal{H}).$ Observable quantities can be associated to Hermitian operators on $\Hi,$ denoted by $\mathfrak{H}(\Hi).$ The {\em support} of an Hermitean operator is the orthogonal complement to its kernel. Given a state $\rho$ and an observable $X$, the expectation of $X$ according to $\rho$ is computed as $\trace(\rho X).$

For any operator $X \in \mathfrak{B}(\mathcal{H})$, denote by $X^{\otimes m}$ the tensor product $X \otimes X \otimes ... \otimes X$ with $m$ factors. Given an operator $\sigma\in{\mathfrak B}(\Hi)$, denote by $\sigma^{(i)}$ the local operator:
\[\sigma^{(i)}:=I^{\otimes (i-1)}\otimes \sigma \otimes I^{\otimes (m-i)}.\]
Permutations of quantum subsystems are expressed by a unitary operator $U_{\pi} \in {\mathfrak U}(\Hi)$, which is uniquely defined by 
\[U_{\pi} (X_1\otimes\ldots \otimes X_m) U_{\pi}^\dag= X_{\pi(1)}\otimes\ldots \otimes X_{\pi(m)}\]
for any operators $X_1,\ldots X_m$ in ${\mathfrak B}(\Hi)$, where $\pi$ is a permutation of the first $m$ integers. ${\mathfrak P}$ is the set of such $\pi$.  A state or observable is said to be \emph{permutation invariant} if it commutes with all the subsystem permutations. 

In \cite{mazzarella-qconsensus} a number of potential extensions of the idea of classical consensus to a quantum network were proposed, and their merit discussed in depth. The ones relevant to this work are:


\begin{defin}[SSC] A state $\rho\in{\mathfrak D}(\Hi^m)$ is in {\em Symmetric State Consensus} (SSC) if, for each unitary permutation $U_{\pi}$,
\[U_{\pi}\, \rho\, U_{\pi}^\dag = \rho \, .\]
\end{defin}

\begin{defin}[$\sigma$SMC]\label{SMC} Given $\sigma\in{\mathfrak B}(\Hi)$ with spectral decomposition $\sigma=\sum_{j=1}^{d}\, s_j \Pi_j \, \in{\mathfrak H}(\Hi)$,  a state $\rho\in{\mathfrak D}(\Hi^m)$ is in {\em Single $\sigma$-Measurement Consensus} ($\sigma$SMC) if:
\beq \trace(\Pi_j^{(k)}\Pi_j^{(\ell)}\rho)=\trace(\Pi_j^{(\ell)}\rho),\label{SMCeq}\eeq
for all $k,\ell\in\{1,\ldots,m\},$ and for each $j$.
\end{defin}

The definition of $\sigma$SMC is the unique, among those proposed, that requires that the outcomes of $\sigma$ measurements on different subsystems be exactly the same {\em for each trial}. 
Consider the set of projections $\{\Pi_j\}^d_{j=1}$ as in Definition \ref{SMC}, and let us define  \[\Pi_{\rm SMC}=\sum_{j=1}^d \Pi_j^{\;\otimes m}.\]
It has been shown that a state is in $\sigma$SMC if and only if it holds
\beq\label{symprop} \trace(\Pi_{\rm SMC}\rho)=1,\eeq or equivalently \beq \Pi_{\rm SMC}\rho\Pi_{\rm SMC}  =\Pi_{\rm SMC}\rho= \rho. \eeq
Furthermore $\sigma$SMC for $\sigma$ with non-degenerate spectrum implies SSC, while the converse implications do not hold.
Lastly, it is impossible for a state to be $\sigma$SMC with respect to all $\sigma\in{\mathfrak H}({\cal H})$: this means that $\sigma$SMC cannot be strengthened to a single-measurement equivalent of SSC.
The proofs of these statements are given in  \cite{mazzarella-qconsensus}.  

It is worth remarking how all these definitions could be given for classical systems, in the context of consensus for random variables or for probability distributions of the state values. In some sense the definition of $\sigma$SMC is the closest to the classical case, as it requires perfect agreement on the outcome of a set of random variables for {\em each measurement}. However, in contrast with the classical ``perfectly correlated'' random variables, the quantum version allows for the output to be determined only at the moment of the measurement. Both SSC and SMC (with non-degenerate observable $\sigma$) imply that the final state is {\em permutation invariant}.

\subsection{Quantum Average Consensus}

So far, only ``static'' properties of consensus states have been discussed. However, the core of the problem of interest for this paper is the design of {\color{black}discrete-time dynamical systems} (or, from an information processing perspective, algorithms) that drive the system to consensus. In addition to this, as in the classical consensus problems, the final state will be required to preserve or express some property dependent on the initial state.

Unitary dynamics are not enough when ones is interested in studying or engineering convergence features for a quantum system. A more general framework that includes (Markovian) open-system evolutions is offered by {\em quantum channels} \cite{kraus,nielsen-chuang}, that is, linear, completely positive (CP) and trace preserving (TP) maps from density operators to density operators $\Ei: \mathfrak{D}(\Hi^m)\rightarrow \mathfrak{D}(\Hi^m).$  It can be shown that such maps admit an {\itshape operator sum representation} (OSR), also known as {\itshape Kraus decomposition}:
\begin{equation}
\label{eq:Kraus}
\Ei(\rho)=\sum_{k=1}^{K}\, A_k\rho A_k^{\dagger} \quad \text{with} \quad \sum_{k=1}^{K} A_k^{\dagger}A_k= I
\end{equation}
where $K\leqslant (\dim(\Hi))^2$.
{The representation is not unique, however the relation between all the possible different representations is well known (see \cite[Theorem 8.2]{nielsen-chuang}). }
A CPTP map is said {\itshape unital} if $\Ei( I)= I$.
Given a CPTP map $\Ei$, its dual map with respect to the Hilbert-Schmidt inner product $\Ei^{\dagger}$ is defined by:
\begin{equation}\label{eq:State/Meas-Duality}
\trace[A\,\Ei(\rho)]=\trace[\Ei^{\dagger}(A)\,\rho] \, .
\end{equation}
This dual map is still linear and completely positive, while the fact that ${\cal E}$ is trace preserving implies that ${\cal E}^\dag $ is unital. 

Locality notions for the quantum network can be introduced as it follows \cite{ticozzi-QL}. Consider the multipartite system introduced in Section \ref{sec:consensus}. An operator $V$ is a {\em neighbourhood operator} with respect to a set of neighborhoods $\{ \mathcal{N}_j \; , \, j=1,2,...,M \}$, if there exists $j \in \{1,2,...,M\}$ such that:
\begin{equation}
\label{eq:quasilocaloperator}
V=V_{\mathcal{N}_j}\otimes I_{\compl{{\mathcal{N}}_j}}
\end{equation}
where  $V_{\mathcal{N}_j}$ accounts for the nontrivial action on $\Hi_{\mathcal{N}_j}$ and $I_{\compl{\mathcal{N}_j}}=\bigotimes_{k\notin \mathcal{N}_j}I_{k}$.

Everything is in place to specify the {\em consensus problems} of interest.  In addition to asymptotically achieving a state with the prescribed ``informational" symmetry, this state must maintain locally some information on the global initial state.
Let $d(\rho_a,{\cal C})=\inf_{\rho \in{\cal C}}\|\rho_a-\rho\|,$ where ${\cal C} \subset {\mathfrak D}(\Hi)$ and $\| \cdot \|$ is any $p$-norm on ${\mathfrak B}(\Hi).$ Given a sequence of QL channels $\{\Ei_t(\cdot)\}_{t=0}^\infty,$ define $\hat{\Ei}_t(\rho_0)=\rho_t={\cal E}_t\circ\Ei_{t-1}\circ\cdots\circ{\cal E}_1(\rho_0),$ and ${\cal C}_{SSC}$ to be the set of states in SSC consensus, and similarly for $\sigma$SMC. Let $S$ be an observable in ${\mathfrak B}(\Hi^{\otimes m}).$
\begin{defin}[Asymptotic $S$-Average Consensus]\label{def:asy}
A sequence of channels $\{\Ei_t(\cdot)\}_{t=0}^\infty,$ is said to \emph{asymptotically achieve SSC} if 
\begin{equation}
\label{evolution}
\lim_{t \rightarrow \infty}d(\hat{\Ei}_t(\rho_0),\mathcal{C}_{\rm SSC})=0,
\end{equation}
for all initial states $\rho_0,$ and {\em there exists} a $\sigma \in {\mathfrak H}(\Hi)$ such that:
\beqa\label{eq:avgconsensus} \lim_{t \rightarrow \infty}\trace(\sigma^{(\ell)}\rho(t))&=&\lim_{t \rightarrow \infty}\trace(S \rho(t))=\trace(S \rho_0)\eeqa for all $\rho_0$ and for all $\ell\in\{1,\ldots ,m\}$. \newline
\end{defin}
The same definition holds for $\sigma$SMC by substituting the corresponding state sets in $(\ref{evolution})$.

\subsection{Gossip-type Dynamics}

Let us recall the structure and core results for the gossip-type algorithm introduced in \cite{mazzarella-qconsensus}. In a controlled quantum network, one can typically engineer unitary transformations that implement the ``identity" evolution and the swapping of two neighboring subsystem states; let us denote the latter operator by $U_{(j,k)}$ for swapping subsystems $j$ and $k$. To develop our analysis, it will be convenient to introduce the {\em graph ${\cal G}$ associated to the multipartite system}: its nodes $1,\ldots,m$ correspond to the ``physical'' subsystems, the edge $(j,k)$ is included if the subsystems $j$ and $k$ (have a non-zero probability to) interact.

{\color{black} Assume edge $(j,k)$ is selected at a certain step $t$: then implement on subsystems $j,k$ the {\em quantum gossip interaction}: 
\begin{equation}\label{alg:gossip}
	\rho(t+1) = {\cal E}_{j,k}(\rho(t))=(1-\alpha)\, \rho(t) + \alpha \, U_{(j,k)}\rho(t)U_{(j,k)}^\dag  \; ,
\end{equation}
with $\alpha \in (0,1)$. 

It has been proven that, if the graph associated to possible interactions is connected, then the quantum gossip algorithm \eqref{alg:gossip} ensures global convergence towards:
\begin{equation}
\label{asymptoLS}
\rho_* = \frac{1}{m!}\, \sum_{\pi \in \mathfrak{P}} \, U_{\pi}\rho_0U_{\pi}^{\dagger}\,\in \mathcal{C}_{SSC}.
\end{equation}
\newline Convergence is {\em deterministic}, when the edges on which a gossip interaction occurs at a given time are selected by periodically cycling, in any predefined way, through the set of available ones; or {\em with probability one\footnote{This means that, for any $\delta,\varepsilon > 0$, there exists a time $T>0$ such that
$$\mathbb{P}[\, \trace((\rho(T)-\rho_*)^2) \,  > \varepsilon \,] \; < \delta \, .$$},} and thus in expectation, when the edges on which a gossip interaction occurs at a given time are selected randomly from a fixed probability distribution $\{ q_{j,k} > 0 \vert \sum_{(j,k)\in E} q_{j,k} = 1 \}$;
Furthermore, $S$-average SSC is attained if and only if $S\in\mathfrak{H}(\Hi^{\otimes m})$ can be written, for some $\sigma\in{\mathfrak H}(\Hi),$ in the form:
\begin{equation}
\label{sumsim}
S=\frac{1}{m}\sum_{i=1}^m\sigma^{(i)}.
\end{equation}
This shows that the mean value of a (global) observable can be asymptotically retrieved from the state of any single subsystem by employing a quantum gossip-type algorithms.

\section{New Results}\label{sec:results}

\subsection{Preliminary observations}

The dynamics associated to the maps \eqref{alg:gossip} are always {\em unital}, that is the identity is always a fixed point for the dynamics. Given the (trace-norm) contraction character of CPTP evolutions \cite{alicki-lendi}, even if the state asymptotically converges  to a different state, the evolution is bound to get closer to the identity, i.e. the completely mixed state.

This type of symmetrization, albeit interesting and satisfying the requisites for asymptotic average SSC, is purity decreasing and in some sense makes the system state ``more classical''. The natural question is then: can quantum consensus be achieved with a pure state?

This, in general, is not possible. The desired dynamics should select a {\em pure} state for any output, while preserving the expectation value of the desired global observable: it would entail a linear map ${\cal E}$ such that:
\beq\trace(S {\cal E}(\rho)) = \trace(S \rho),\label{eq:preservedS}\eeq
and $\trace{\cal E}(\rho)^2)=1.$
Write $S$ in spectral decomposition as $S=\sum_k\lambda_k\Pi_k,$ with $\Pi_k$ the orthogonal projectors onto the (degenerate) eigenspaces ${\cal V}_k$ of $S$. For any pure state $|\psi_k\rangle\in{\cal V}_k,$ the output ${\cal E}(|\psi_k\rangle\langle \psi_k |)=|\phi_k\rangle\langle\phi_k|$ should remain pure. However, any linear map ${\cal E}$ would satisfy
\beqan{\cal E}(\lambda_k|\psi_k\rangle\langle \psi_k|&+&(1-\lambda)|\psi_j\rangle\langle \psi_j|)\\
&&=\lambda_k{\cal E}(|\psi_k\rangle\langle \psi_k|)+(1-\lambda){\cal E}(|\psi_j\rangle\langle \psi_j|)\\
&&=\lambda_k|\phi_k\rangle\langle\phi_k|+(1-\lambda_j)|\phi_j\rangle\langle\phi_j|,\eeqan
where $0<\lambda<1.$  The only case in which the output can be pure for any $\lambda$ is when ${\cal E}(\psi_k\rangle\langle \psi_k|)={\cal E}(|\psi_j\rangle\langle \psi_j|):$ this would satisfy the expectation preservation if and only if $|\psi_1\rangle$ and $|\psi_2\rangle$ had the same expectation, which is clearly not possible for $j\neq k$.
Inspired by this observation, we will construct dynamics that select a pure representative in each of the eigenspaces of the global observable $S$, yet losing some of the correlations between subspaces.

In the rest of the paper, in order to limit the notational complexity and provide some matrix representations that are easier to visualize, the focus is on: 
(1) networks of two-level systems or qubits, $\Hi_i=\span\{\ket{0},\ket{1}\}.$ We use the common shorthand notation for the standard basis of a composite Hilbert space $\Hi^m$: $\ket{01\ldots}=\ket{0}\otimes\ket{1}\otimes\ldots$;
(2) pair-wise interactions, i.e. the neighborhoods are a subset of the set of  pairs $\{\{j,k\},\;1\geq j,k\leq m, j\neq k\}.$ 
Both assumptions can be relaxed, with some straightforward adaptation in the algorithms.

\subsection{Symmetrizing dynamics for symmetric state consensus}

The preservation of the asymptotic expectations is clearly guaranteed if each eigenspace of $S$ is an invariant subspace for all maps that can enter the evolution, at any time.
More precisely, if $S$ is written in spectral decomposition as $S=\sum_k\lambda_k\Pi_k,$ with $\Pi_k$ the orthogonal projectors onto the (degenerate) eigenspaces of $S$, call them $\Hi_{k}$, invariance of all the $\Hi_k$ is equivalent to require ${\cal E}^\dag(\Pi_k)=\Pi_k$ \cite{cirillo-decomposition}, so that:
\[\trace({\cal E}_{t}(S) \rho)=\trace(\sum_k\lambda_k{\cal E}_{t}(\Pi_k) \rho)=\trace(S \rho).\]
On the other hand, as long as convergence to a consensus state is guaranteed, dynamics inside the subspace is arbitrary.
Let us build one that, while satisfying eigenspace invariance, prepares a pure representative in each $S$ eigenspace.
For networks of two-level systems, up to a local change of basis and a rescaling, any $S$ as in \eqref{sumsim} can be rewritten as 
\begin{equation}
\label{sumsim1}
S=mI+\sum_{i=1}^m\sigma_z^{(i)}.
\end{equation}
This allows to identify the eigenspaces of $S$ with respect to this basis as the ``excitation'' subspaces:
\beq\label{eq:seigs} \Hi_k=\span\{\ket{ i_1,i_2,\ldots, i_m},i_k\in\{0,1\}, \sum_{\ell=1}^m i_\ell=k\}
\eeq
generated by linear combination of vectors belonging to the standard basis associated to exactly $k$ $1$'s. As typical in the physics literature, call the number of 1's in such vectors, and thus of the generated subspaces, the {\em excitation number}.
Each of these subspaces contains {\em only one} pure symmetric state, which is referred in the physics literature as a {\em Dicke} state \cite{dicke,dicke-rev}.
Each state is labeled by the number of systems $N$ and the number of excitations $k$, and is associated to a vector:
\begin{equation}
 \ket{(m,k)}=\frac{1}{\sqrt{{m \choose k}}}(\ket{\underbrace{1\ldots 1}_k\underbrace{0\ldots 0}_{m-k}}+\ldots+\ket{\underbrace{0\ldots 0}_{m-k}\underbrace{1\ldots 1}_k}),
\end{equation}
and each term of the sum corresponds to a unique arrangement of the $k$ ones and $N-k$ zeros.
To our aim, the key property of Dicke states, beside being pure and symmetric, is that they admit a very simple {\em Schmidt decomposition.} Consider a partition of the network into two groups of $m_A$ and $m_B$ subsystems, respectively. It is then easy to show that the Schmidt decomposition of a Dicke state $\ket{(m,k)}$} is
\begin{equation}\label{eq:decomposition}
 \ket{(m,k)}=\frac{1}{\sqrt{{m \choose {{k}}}}}\sum_{{k}_A+{k}_B=k}\mu_{k_A,k_B}\ket{(m_A,k_A)}\otimes\ket{(m_B,k_B)},
\end{equation}
where the weight on each term is $\mu_{k_A,k_B}=\sqrt{{m_A \choose k_A}{m_B \choose k_B}}$.

In order to explicitly construct the dynamics that prepare these states in each eigenspace, choose a neighbourhood ${\cal N}_j,$ 
containing $2$ qubits, and consider the neighbourhood operator:
\begin{equation}
S_{{\cal N}_j}=2I+\sum_{i\in{\cal N}_j}\sigma_z^{(i)}.
\end{equation}
As we did for the global $S$ in \eqref{eq:seigs}, we can construct the eigenspaces $S_{{\cal N}_j}$ as the subspaces spanned by pure states with a given number of excitation:
\beq \label{eq:seigs1}\Hi_k^{{\cal N}_j}=\span\{\ket{ i_1,i_2,\ldots, i_{m_j}},i_\ell\in\{0,1\}, \sum_{\ell=1}^{m_j} i_\ell=k,\}\eeq
which, in our two-qubit neighborhood, are just three:
\[\Hi_0^{{\cal N}_j}=\span\{\ket{ 00}\},\]
\[\Hi_1^{{\cal N}_j}=\span\{(\ket{ 01}+\ket{ 10})/\sqrt{2},(\ket{ 01}-\ket{ 10})/\sqrt{2}\},\]
\beq\Hi_2^{{\cal N}_j}=\span\{\ket{ 1,1}\}.\eeq
Joining these sets of vectors (in order), a basis for $\Hi_0^{{\cal N}_j}\oplus\Hi_1^{{\cal N}_j}\oplus\Hi_2^{{\cal N}_j}$ is obtained. With respect to this basis, construct a neighbourhood CPTP map $\tilde \Ei_{\Ni_j}$ with operators:
\beq \tilde M_{j,1}=\left[\begin{array}{c|cc|c}
0 & 0 & 0 & 0\\
\hline 0 & 0 & 1 & 0\\
0 & 0 & 0 & 0 \\ 
\hline 0 & 0 & 0 & 0 \\ 
\end{array}\right]\eeq

\beq \tilde M_{j,2}=
\left[\begin{array}{c|cc|c}
1 & 0 & 0 & 0\\
\hline 0 & 1 & 0 & 0\\
0 & 0 & 0 & 0 \\ 
\hline 0 & 0 & 0 & 1 \\ 
\end{array}\right].\eeq
%
\noindent From these neighborhood operators, it is easy to construct a map on the whole network, which is going to be the dynamics associated to the selection of neighborhood ${\cal N}_j$:
\beqa \label{eq:iteration} \Ei_{{\cal N}_j}(\rho)&=&\tilde\Ei_{{\cal N}_j}\otimes {\cal I}_{\bar{\cal N}_j}(\rho)\\
&=&\tilde M_{j,1}\otimes I\,\rho\,\tilde M_{j,1}^\dag\otimes I+\tilde M_{j,2}\otimes I\,\rho\,\tilde M_{j,2}^\dag\otimes I.\nonumber\eeqa
Here ${\cal I}_{\bar{\cal N}_j}$ is the identity map on $\mathfrak{B}(\Hi_{\bar{\cal N}_j}),$ which is the set of operators that are forced to evolve trivially for neighborhood maps on ${\cal N}_j.$ 
\noindent Now notice that each map is CPTP, and:
\begin{enumerate}
\item The first operator is nilpotent and acts like a generalized ``raising'' operator, which ensures that invariant {\em neighbourhood} states for the map $\tilde\Ei_{{\cal N}_j}$ must have support on 
\[\Hi^0_j=\span\{\ket{(2,\ell)},\ell=0,1,2\},\]
which is the kernel of $\tilde M_{1,j}.$ 
\item The second operator $\tilde M_{j,2}$ is just the orthogonal projector onto $\Hi^0_j.$ The latter is thus an invariant subspace for $\tilde\Ei_{{\cal N}_j}$, and any state with support on it is a fixed point. \item Given the ladder structure of $\tilde M_{1,j},$ any state for the neighborhood that does not have support on it is attracted to $\Hi^0_j.$
\end{enumerate}

In terms of typical available gates/control resources for pairs of qubits, such a map can be implemented in a ``digital quantum simulator'' \cite{quantumsimulator}, with a two-body unitary (that e.g. maps the ordered standard basis into $\{(\ket{0}+\ket{1})/\sqrt{2},(\ket{0}-\ket{1})/\sqrt{2},\ket{00},\ket{11}\}$), a single qubit reset and another two body unitary reverting the effect of the first. These maps can also be implemented via discrete quantum feedback \cite{bolognani-arxiv}, via a direct projective measurement with two outcomes, associated to projectors $\{\Pi_1=I-\tilde M_{j,2},\Pi_2=\tilde M_{j,2}\},$ followed by a unitary operation when the outcome associated to the first projector is observed:
\[\tilde M_{j,1}=U_1\Pi_1,\quad U_1=\left[\begin{array}{c|cc|c}
0 & 0 & 0 & 0\\
\hline 0 & 0 & 1 & 0\\
0 & 1 & 0 & 0 \\ 
\hline 0 & 0 & 0 & 0 \\ 
\end{array}\right],\]
where the matrix representation is given with respect to the basis constructed above.

All is in place to prove the following:
\begin{thm}
\label{thm:convergencenew}
If the neighborhoods cover the whole networks and the associated graph\footnote{The graph that has subsystems as nodes, and edges connecting pairs of subsystems that belong to same neighborhood.} is connected, global convergence towards SSC is ensured:
\newline - {\em Deterministically}, when all the neighborhoods in any order such that all neighborhoods are selected an infinite number of times;
\newline - {\em In probability and in expectation}, when all the neighborhoods are selected  randomly from a fixed probability distribution $\{ q_j=\Prob[{{\cal N}_j}] > 0 \vert \sum_{j} q_{j} = 1 \}$.
In any of the above cases, the system converges to a state $\rho_*$ that has support on 
\beq \Hi^0=\span\{\ket{(m,\ell)},\ell=0,\ldots,m\}.\eeq
Furthermore, if the initial state has support on the global eigenspaces of $S$ with eigenvalue $\lambda,$ the final state is the pure state $\rho_*=\ket{(m,\lambda)}\bra{(m,\lambda)}.$
\end{thm}
{\proof}
Let $\Hi^0$ be the span of  the Dicke states on the {whole network}. From decomposition \eqref{eq:decomposition}, it follows that for each $j,$ it holds that: \beq\Hi^0\subseteq \Hi^0_j\otimes \Hi^0_{\bar{\cal N}_j}.\eeq
Hence $\Hi^0\subseteq \bigcap_j\Hi^0_j\otimes \Hi^0_{\bar{\cal N}_j}$ and it is invariant for each map as well.
It is easy to see that $\Hi^0$ is actually $\Hi^0=\bigcap_j\Hi^0_j\otimes \Hi^0_{\bar{\cal N}_j}.$ In fact, if a vector is not in $\Hi^0,$ then it is also not in $\Hi^0_j$ for some $j.$ If a state $\rho$ is invariant for all maps ${\cal E}_{{\cal N}_j},$ then it has support on the symmetric subspace. 
Since they do not change the excitation number, each $\Ei_{{\cal N}_j}$ leaves the eigenspaces of the corresponding $S_{{\cal N}_j}$ invariant, as well as those of the global $S.$ 

Consider now a finite sequence of $T$ neighborhood maps $\{\Ei_{{\cal N}_{j(t})}\}_{t=1,2,\ldots, T}$ in which enough overlapping neighborhood are chosen so that they cover the whole network of $m$ qubits, and the define the composite map:
\[\Ei^o_m=\Ei_{{\cal N}_{j(T)}}\circ \ldots\circ \Ei_{{\cal N}_{j(1)}}.\]
Since each $\Ei_{{\cal N}_{j(t})}$ leaves the eigenspaces of $S$ invariant, the same is true for $\Ei^o_m.$

Consider the invariant subspaces $\Hi_k$ of $S$, and define $V^m_k(\rho)=1-\bra{(m,k)}\rho\ket{(m,k)},$ and the ($m$-system) Dicke density operators as \beq\rho_{m,k}=\ket{(m,k)}\bra{(m,k)}.\eeq Clearly $V_{k}^m(\rho_{m,k})=0,$ and $V^m_k(\rho)>0$ if $\rho\neq\rho_{m,k},$ and it will be our Lyapunov-type functions: if it is true that for all $k$ and $\rho\neq\rho_{m,k}$ with support in $\Hi_k$ there is a sequence of maps such that:
\beq\Delta V^m_{k}(\rho):=V^m_k({\cal E}^o_{m}(\rho))-V^m_k(\rho)<0,\label{eq:key}\eeq
then the only state with support on $\Hi_k$ such that $\Delta V^m_{k}(\rho)=0$ would be $\rho_{m,k}.$ Hence, considering $V(\rho)=\sum_kV^m_k(\rho),$ the main Theorem statement would follow by the discrete-time version of the {\em invariance principle} (see Appendix \ref{lasalle}), since we would prove convergence to $\Hi^0$, and it would be possible to further conclude that $\rho_{m,k}$ is asymptotically stable in $\mathfrak{D}(\Hi_k)$.

The key is thus to prove that there exists a sequence such that \eqref{eq:key} holds, by induction on the number of qubits $m$. The inductive hypothesis is: {\em  For $m$ qubits and all $0\leq k\leq m$, there exists ${\cal E}^o_{m}$ such that $\Delta V^m_{k}(\rho)<0$ for any $\rho\in\Di(\Hi_{m,k}),$  $\rho\neq \rho_{m,k}.$}\\
\noindent For $m=2,$ the network consists in a single pair, so there is a single neighborhood $\Ni=\{1,2\}$, and   
$\Ei^o_2=\Ei_{{\cal N}}.$ By construction, $\Ei_{{\cal N}}$ satisfies the inductive hypothesis. In fact,
\[\Hi_0^{{\cal N}}=\span\{\ket{ 0,0}\},\Hi_2^{{\cal N}}=\span\{\ket{ 1,1}\}\]
only support the invariant states $\rho_{2,0},\rho_{2,2},$ respectively, while on
$\Hi_1^{{\cal N}_j}$ it prepares the state $\rho_{2,1}=(\ket{ 0,1}+\ket{ 1,0})(\bra{ 0,1}+\bra{ 1,0})/{2}$ with a single application, so $\Delta V^m_{1}(\rho)=-V^m_{1}(\rho)<0$.

\noindent Next, the inductive step is proved.
Choose a partition of the $m+1$ qubits in $2$ qubits and the remaining $m-1$ ones: without loss of generality (up to a relabelling), choose $\Ni_1=\{1,2\}$ and $\overline\Ni_1$. 
Consider a $\Ei^o_{m+1}$ of the form $\Ei_{\Ni_1}\circ\Ei^o_{m}.$ Since $\rho_{m+1,k}$ is invariant for all neighborhood maps, both $V^{m+1}_k({\cal E}^o_{m}(\rho))-V^{m+1}_k(\rho)\leq 0$ and $V^{m+1}_k({\cal E}_{\Ni_{1}}(\rho))-V^{m+1}_k(\rho)\leq 0$ for any initial $\rho$ \cite{bolognani-arxiv}. It is next shown that at least  one of them is strictly negative if $\rho\neq\rho_{m+1,k}.$ If it holds already $V^{m+1}_k({\cal E}^o_{m}(\rho))-V^{m+1}_k(\rho)<0,$ then the induction step is already proved. Recall that $\rho_{m+1,k}$ is the projector onto 
\beqan\span\{\ket{(m+1,k)}\}&=&\span\big\{\ket{0}\otimes\ket{(m,k)}\\&&+\ket{1}\otimes\ket{(m,k-1)}\big\}\\
&=&\span\big\{\ket{00}\otimes\ket{(m-1,k)}\\&&+(\ket{01}+\ket{10})\otimes\ket{(m-1,k-1)}\\&&+\ket{11})\otimes\ket{(m-1,k-2)}\big\},\eeqan
and
\beqan\Hi_{(m+1,k)}&=&\span\big\{\ket{0}\}\otimes\Hi_{(m,k)}\oplus\span\{\ket{1}\big\}\otimes\Hi_{(m,k-1)}.\eeqan
By the inductive hypothesis and conservation of the excitation number $k$, it follows that ${\cal E}^o_{m}(\rho)$ guarantees  $\Delta V^{m+1}_k<0$ for any state on $\Hi_{(m+1,k)},$ unless:
%
\beqan\supp(\rho)&\subseteq&\left(\span\{\ket{0}\otimes\ket{(m,k)}\}\right)\oplus\\&& \left(\span\{\ket{1}\otimes\ket{(m,k-1)}\}\right).
\eeqan
Since $\rho_{m,k}$ is invariant for all neighborhood maps, then considering ${\cal E}^o_{m+1}(\rho)=\Ei_{\Ni_k}\circ{\cal E}^o_{m}(\rho)$ still makes $V_{k}^m$ decrease.\\
Consider now the case in which ${\cal E}^o_{m}$ leaves $V_k^m$ unvaried and let us consider an additional map $\Ei_{\Ni_1},$ with the neighborhood overlapping, but not being included, with the subsystems non-trivially affected by ${\cal E}^o_{m}(\rho).$ In terms of the $\Ni_1,\overline\Ni_1$ partition, this means:
\beqan
&&\supp(\rho)\subseteq\Hi_{\rm inv}\\
&&=\left(\span\{\ket{00}\otimes\ket{(m-2,k)}+\ket{01}\otimes\ket{(m-2,k-1)}\}\right)\\&&\oplus \left(\span\{\ket{10}\otimes\ket{(m-2,k-1)}+\ket{11}\otimes\ket{(m-2,k-2)}\}\right)
\eeqan

The action of $\Ei_{\Ni_1}$ maps the subspace orthogonal to the target in $\Hi_{\rm inv},$ namely the span of:
\beqan\ket{\phi^\perp}=&&\ket{00}\otimes\ket{(m-1,k)}\\&&+(\ket{01}-\ket{10})\otimes\ket{(m-1,k-1)}\\&&-\ket{11}\otimes\ket{(m-1,k-2)}\eeqan
to 
\beqan&&\span{\big\{}\ket{00}\otimes\ket{(m-1,k)}\\&&+(\ket{01}+\ket{10})\otimes\ket{(m-1,k-1)}\\&&-\ket{11}\otimes\ket{(m-1,k-2)}.{\big\}}\eeqan
This implies that either $\rho=\rho_{m+1,k}$ or:
\[
V^{m+1}_k({\cal E}_{\Ni_{1}}(\rho))-V^{m+1}_k(\rho)=-\frac{1}{2}\bra{\phi^\perp}\rho\ket{\phi^\perp}<0.
\]
Notice that is shows it is sufficient to choose the {sequence of neighbourhoods, and thus the sequence of maps, so that they have always overlap with the  already selected  ones and eventually cover the whole network}. This is also shown by induction on $m$: for $m=2$ it is trivially true with $\Ei^o_2$ as no other choices of neighborhoods is possible. If it is true for $m,$ and the graph associated to the neighborhoods is connected,  any sequence of neighborhood that satisfies the above properties on some connected sub-graph of $m$ qubits and adding an additional two-bit neighborhood that covers the $m+1$-th qubit can be used. The proof above applies to this case, and the Lyapunov function strictly decreases for sequences constructed in this way.\\
 Convergence in probability is then proved by a direct application of Lemma \ref{lem:randconv} in the Appendix, using the sequence associated to $\Ei^o_m,$ and convergence in expectation trivially follows from convergence in probability.
\qed

From Theorem \ref{thm:convergencenew}, it is easy to prove the following result.
\begin{cor}\label{cor:prep}
If the following control capabilities are available:\\
(i) Projective measurements of $\sigma_z$ and unitary control $\sigma_x$ for the single qubits;\\
(ii) Pairwise interactions as in \eqref{eq:iteration} on a connected graph;\\
Then any Dicke state is asymptotically preparable. If, in addition, projective measurements of $S$ as in \eqref{sumsim1} are available, any Dicke state can be made asymptotically stable.
\end{cor}
\proof
Consider a target Dicke state, and assume it is in the $\ell$-eigenspace of $S.$ The strategy to prepare it is the following:
\begin{enumerate}
\item First, we want to prepare the correct eigenspace of $S.$ To do so, if this additional resource is available, first measure $S$: if the outcome is $\ell$, move to the next step. Otherwise, if either we cannot measure $S$ or the eigenspace is not the correct one, start measuring the single qubits. Let $\ell'$ be the number of $1$ outcomes. If $\ell-\ell'=q>0$ apply $\sigma_x$ to $q$ qubits that gave $-1$ as outcome, if $q<0$ apply $\sigma_z$ to $q$ qubits that gave $1$ as outcome. This will prepare a pure factorized, yet not symmetric, state in the correct eigenspace.
\item Next, run the SSC algorithm of Theorem \ref{thm:convergencenew} using any cyclic repetition of overlapping pairwise interactions that cover the whole network.
 \end{enumerate}
 By the previous result, we will have convergence to $|(m,\ell)\rangle\langle(m,\ell)|.$
 Notice that if $S$ can be measured and the initial state is already the correct Dicke state, it will remain invariant for the dynamics, as single qubit measurements will not be applied. Hence, the above protocol effectively asymptotically stabilizes the target. On the other hand, if $S$ cannot be measured, the single qubit measurements will initially disrupt the desired state, which will be re-prepared by the second step, as in quasi-local conditional stabilization procedures of \cite{ticozzi-QL}. 
\qed
This can be thought as a discrete-time, generalized version of the strategies proposed to conditionally stabilize W states with continuous semigroup dynamics in \cite{ticozzi-QL}. Dicke states, with the exception of the ones corresponding to extremal eigenspaces, are entangled and of interest for quantum information applications \cite{nielsen-chuang}.

\subsection{Symmetrizing Dynamics for Single-Measurement Consensus}

The following dynamics addresses average $\sigma$SMC. Given Theorem 1, it is necessary to design QL dynamics that drive the state on the support of the projector $\Pi_{\rm SMC},$ while maintaining the expected value of $S.$ In the qubit network case, after putting $S$ in the standard form \eqref{sumsim1}, the support of $\Pi_{\rm SMC}$ corresponds to:
\[\Hi_{\rm SMC}=\span\{|00\ldots 0\rangle,|11\ldots \rangle\}.\]

In order to construct the stabilizing neighborhood maps, consider the standard basis $\{\ket{k_j}\}_{k=0}^{d_{{\cal N}_{j}-1}}$ for $\Hi_{{\cal N}_{j}},$ where $\ket{0}=\ket{00\ldots 0},$ $\ket{1}=\ket{00\ldots 01},$ and so on with binary ordering until $\ket{d_{{\cal N}_{j}-1}}=\ket{11\ldots 1}.$

Define $\Pi_{{\cal N}_{j},k}=\ket{k_j}\bra{k_j}$ and  $\Pi_{{\cal N}_{j},\rm sym}= \Pi_{{\cal N}_{j},1}+\Pi_{\Ni_j,d_{{\cal N}_{j}-1}}.$ This is a projector on the neighborhood reduction of $\Hi_{\rm SMC},$ namely $\Hi_{\Ni_j,{\rm SMC}}. $ 
 Now construct a neighbourhood CPTP map:
\beqa \label{eq:SMCstr} \tilde\Ei_{{\cal N}_{j}}(\rho)&=& \Pi_{{\cal N}_{j},\rm sym}\rho\Pi_{{\cal N}_{j},\rm sym} + \sum_{k=1}^{d_{{\cal N}_{j}-2}} \tilde{\cal R}_{{\cal N}_{j},k}(\Pi_k\rho\Pi_k),\eeqa
where:
\beqa
\tilde{\cal R}_{{\cal N}_{j},k}(\tau)&=&\sum_{k=1}^{d_{{\cal N}_{j}-2}} \left(p_{k,0}U_{k,0}\tau U_{k,0}^\dag + p_{k,1}U_{k,1} \tau U_{k,1}^\dag\right),\nonumber
\eeqa
and: 
\begin{enumerate}
\item $U_{k,0}$ satisfies $U_{k,0}\ket{k}=\ket{0}$ and  $U_{k,1}$ satisfies $U_{k,1}\ket{k}=\ket{d_{{\cal N}_{j}-2}}$;  
\item $p_{k,0}$ is equal to the number of zeros in the binary representation of $k$ divided by the number of qubits in the neighborhood, and $p_{k,1}=1-p_{k,0}$.\end{enumerate} 
Intuitively, $\tilde\Ei_{\Ni_j}$ leaves any state with support on $\Hi_{\Ni_j,{\rm SMC}}$ invariant, while on the orthogonal it performs a measurement projecting the state on the standard basis for the neighborhood, followed by a ``randomizing'' map $\tilde {\cal R}_{{\cal N}_{j},k}.$ This map transfers each basis vector to a mixture of $|00\ldots 0\rangle,|11\ldots 1\rangle, $ with weights $p_{k,0},p_{k,1}$ chosen so that they preserve the expectation of $S$.  

The maps on the whole network are obtained by the neighborhood maps as:
\beq \label{alg:SMC} \Ei_{{\cal N}_j}(\rho)=\tilde\Ei_{{\cal N}_j}\otimes {I}_{\bar{\cal N}_j}(\rho)\eeq
By construction, each of these maps is CPTP, and leave any state on $\Hi_{\rm SMC}$ invariant. The following convergence result holds:

\begin{thm}
\label{thm:convergencenew}
If the neighbourhoods cover the whole networks and the associated graph is connected, then the dynamics that applies  \eqref{alg:SMC} when the neighbourhood ${\cal N}_j$ is selected ensures global convergence towards $\sigma$SMC:
\newline - {\em Deterministically}, when all the neighborhoods are selected cyclically;
\newline - {\em In probability and in expectation}, when all the neighborhoods are selected  randomly from a fixed probability distribution $\{ q_j=\Prob[{{\cal N}_j}] > 0 \vert \sum_{j} q_{j} = 1 \},$ independently at each time.
\end{thm}
\proof
The density operators with support on $\Hi_{\rm SMC}$ are the unique common fixed states for all the $ \Ei_{{\cal N}_j}$ maps: if  state has support on some subspace orthogonal to $\Hi_{\rm SMC},$ there is at least a neighborhood $\Ni_j$ in which the reduced state has support outside $\Hi_{\Ni_j,{\rm SMC}},$ so it cannot be invariant for the corresponding $\Ei_{\Ni_j}.$ 
Consider 
\[V_m(\rho) =  1 - \trace(\Pi_{{\rm SMC},m}\rho),\]
where $ \Pi_{{\rm SMC},m}$ is the orthogonal projection on $\Hi_{\rm SMC}$ for $m$ qubits.
It shall be proved by induction on the number of subsystems $m$ that there exists a subsequence of $T$ maps:
\[\Ei^o_m=\Ei_{{\cal N}_{j(T)}}\circ \ldots\circ \Ei_{{\cal N}_{j(1)}},\]
such that for any $\rho$ with support outside of $\Hi_{\rm SMC}$:
\[\Delta V_m(\rho):=V({\cal E}^o_{m}(\rho))-V(\rho)<0.\]
For $m=2$ and ${\cal E}^o_{2}=\Ei_{\{1,2\}},$ the result is immediate given the structure \eqref{eq:SMCstr}: if a state has support outside of $\Hi_{rm SMC},$ it must have support on $\span\{|01\rangle,|10 \rangle\}.$ Assume that $q=\langle 01|\rho|01\rangle + \langle 10|\rho|10\rangle>0.$ Then the effect of the randomizing feedback map is that of transporting the state inside $\Hi_{\rm SMC},$ so that:
\[\Delta V_m(\rho):=V_m({\cal E}^o_{2}(\rho))-V_m(\rho)=-q<0.\]
Now assume that the inductive hypothesis holds for $m$ subsystems, with some sequence of neighborhood associated to a map ${\cal E}^o_{m}.$ If another subsystem is added, the effect of ${\cal E}^o_{m}$ on this $m+1$ multipartite system is evaluated.
By the inductive hypothesis: 
\[\Delta V_m(\rho):=V_m({\cal E}^o_{m}(\rho))-V_m(\rho)<q<0.\]
This means that the trace of the output state onto $\Hi_{{\rm SMC},m}$ has increased. The subspace $\Hi_{{\rm SMC},m}\otimes \Hi_{m+1}$ has thus also gain probability. This subspace is spanned by $|00\ldots 0\rangle\otimes|x\rangle,
|11\ldots 1\rangle\otimes|y\rangle,$ with $x,y\in\{0,1\}.$ If the map $\Ei_{\{m,m+1\}}$ is applied, anything in this subspace is mapped onto $\Hi_{{\rm SMC},m+1}.$ Thus it also holds that
\[\Delta V_{m+1}(\rho):=V_{m+1}({\cal E}^o_{m+1}(\rho))-V_{m+1}(\rho)<q<0.\]
In this way it has been proved that a sequence of maps that contracts $V_m$ for any $m$ does exists.
By considering cyclic application of the sequence associated to the map ${\cal E}^o_{m+1}$, a straightforward application of the {\em invariance principle} (in the discrete-time version \cite{lasalle-discrete}), asymptotic convergence is achieved.\\ Convergence in probability is then proved via Lemma \ref{lem:randconv} in the Appendix, using the sequence associated to $\Ei^o_m,$ and convergence in expectation directly from convergence in probability.
\qed
The probability distribution over the neighborhoods can actually be made time dependent, as long as all of its elements stay bounded away from zero (see e.g. the proofs of convergence of \cite{mazzarella-group}).

\section{A Toy Example}\label{sec:example}

\begin{figure}[!h]
  \hspace{-3mm}\includegraphics[width=9.5cm]{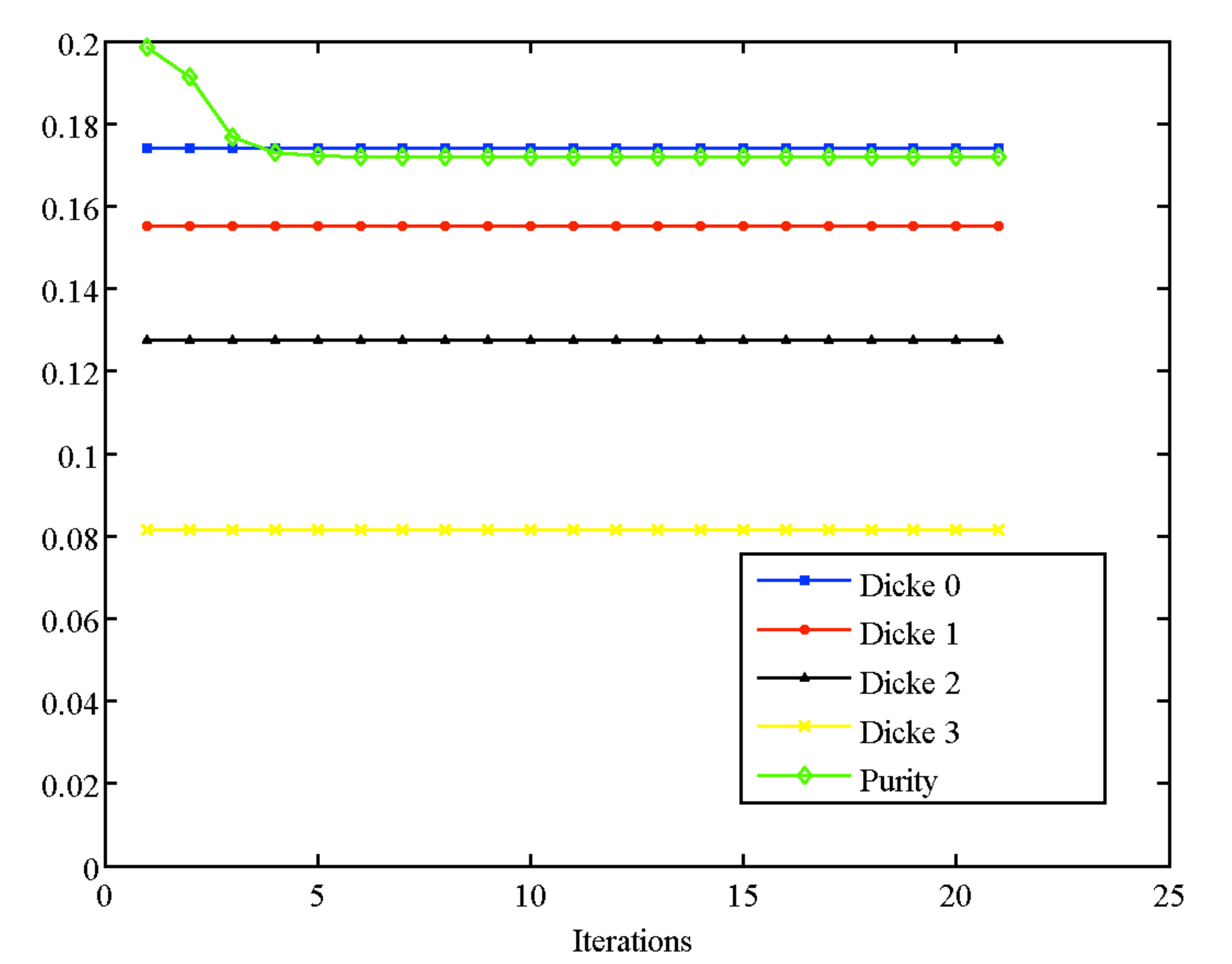}
  \caption{(Color online) {Purity and population dynamics for the Dicke states induced by the gossip-type algorithm GOS.}}
  \label{fig:GOS}
\end{figure}

\begin{figure}[!h]
   \hspace{-3mm}\includegraphics[width=9.5cm]{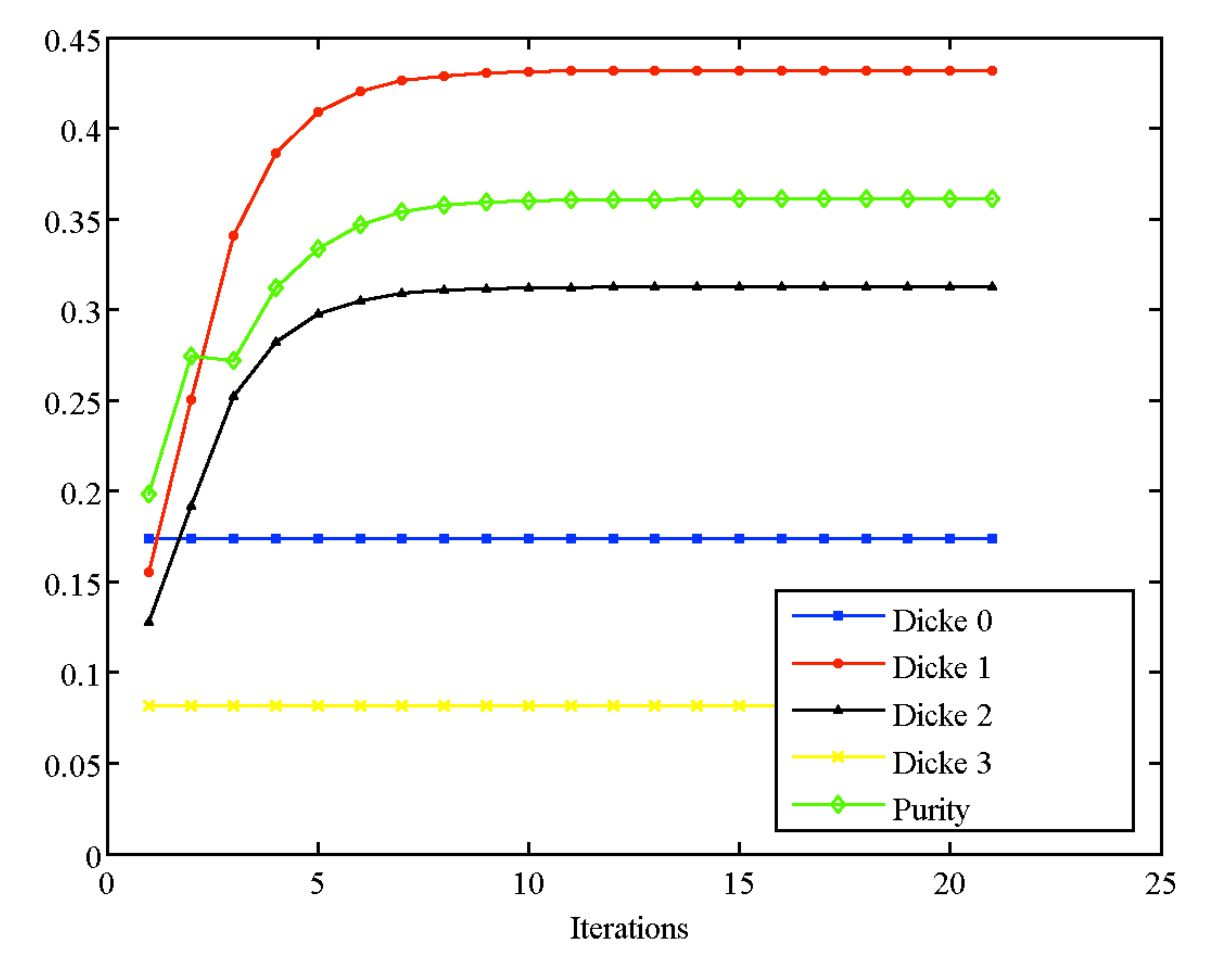}
  \caption{(Color online) {Purity and population dynamics for the Dicke states induced by the SSC algorithm}}
  \label{fig:SSC}
\end{figure}

\begin{figure}[!h]
   \hspace{-3mm}\includegraphics[width=9.5cm]{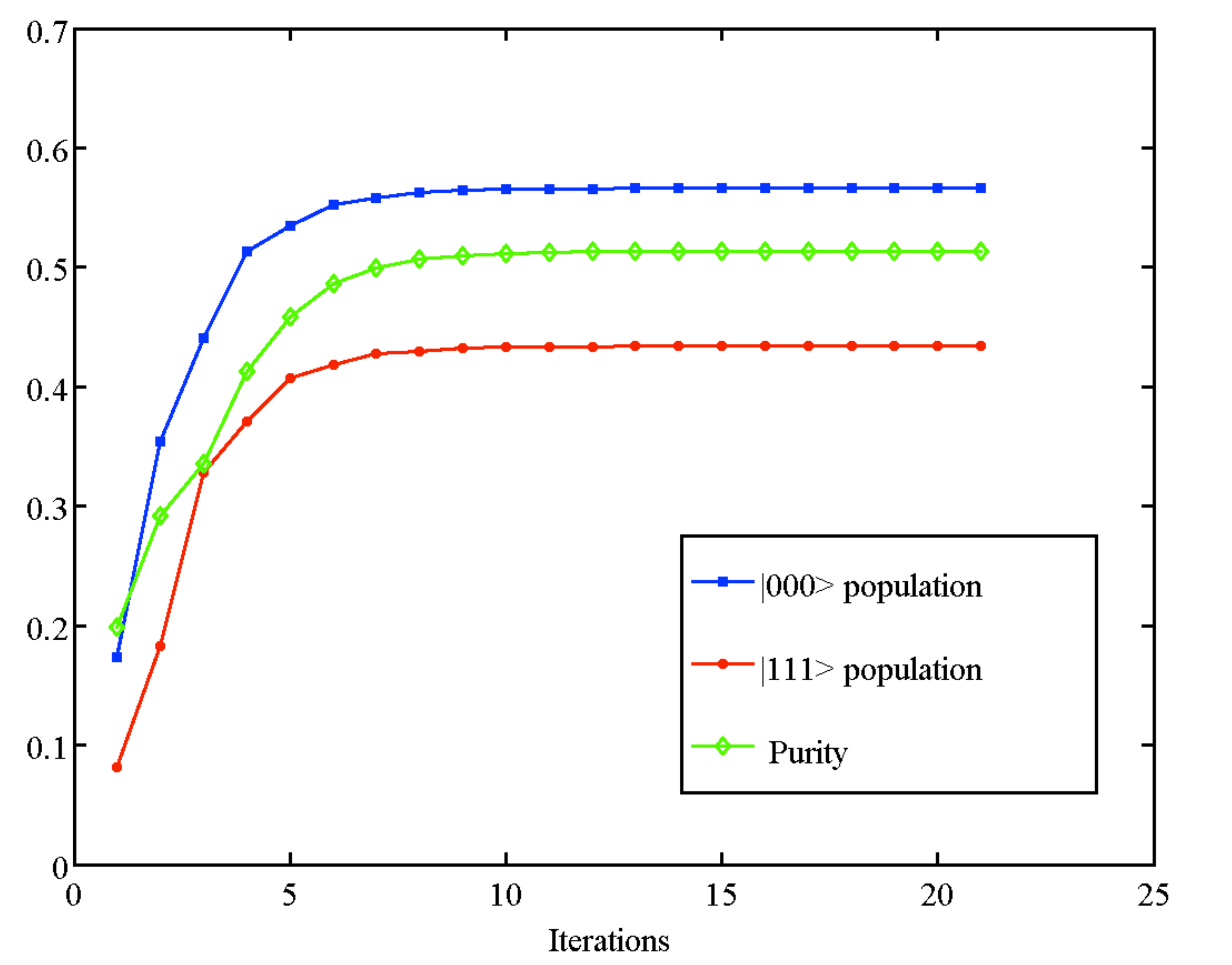}
  \caption{(Color online) {Purity and population dynamics for the $\ket{000},\ket{111}$ states induced by the SMC algorithms.}}
  \label{fig:SMC}
\end{figure}

Consider three systems of dimension two, labelled as $1,2,3,$ and allow for controlled interactions on neighborhoods $\{1,2\}$ and $\{2,3\}$. The three algorithms for symmetrization that are compared, identified by the acronyms below, are respectively using:
\begin{description}
\item[(GOS):] The gossip-type interactions \eqref{alg:gossip}; 
\item[(DSC):] The Dicke-preparing interactions \eqref{eq:iteration};
\item[(SMC):] The SMC-preparing interactions \eqref{alg:SMC}.
\end{description}
For all the above, alternating actions on neighborhoods $\Ni_1=\{1,2\},\Ni_2=\{2,3\},$ are considered. The presented simulations illustrate the different behavior of the dynamics proposed in the preceding sections.  An initial state is generated by randomly selecting its spectrum and then randomly choosing a unitary orthonormal basis in which it is diagonal, and the evolution of the purity, along with some relevant state populations, are plotted as in Figures \ref{fig:GOS}-\ref{fig:SMC}.

 The main observations are summarized in the following.
 \begin{itemize}
 \item Given the theoretical analysis of the previous sections, the final state for the SSC algorithm has support only on the span of the four symmetric Dicke vectors, and thus the evolution of the four respective populations is depicted in the figure. Similarly for the SMC algorithm and the $\ket{000},\ket{111}$ vectors.
 \item Figure \ref{fig:SSC} shows how the average SSC-attaining algorithm associated to maps of the form \eqref{eq:iteration} saturates the population of the Dicke states, overall improving the purity of the state. It is worth noting that the increase is not monotone. The natural comparison is with the gossip-type algorithm \eqref{alg:gossip}, which also attains average SSC: as illustrated in Figure \ref{fig:GOS}, by projecting the state on the group commutant \cite{mazzarella-qconsensus,mazzarella-group} with unital operations, the purity decreases while the populations of the Dicke states are invariant. 
While the above behavior is typical for random initial states, if the initial states is e.g. pure its purity will not typically be preserved even for the new SSC algorithm, unless it is a state that belongs to one of the eigensubspaces of $S.$
 \item The SMC algorithm symmetrizes the state by ``pushing'' it in the subspace generated by $\ket{000},\ket{111}.$ The increase of the population of these subspaces is depicted in Figure \ref{fig:SMC}, along with the associated improvement in purity. A direct comparison with the other algorithms shows that, for the given initial state, this method leads to the highest final purity, with a final value of $0.51$ for SMC, versus $0.36$ for the SSC and $0.17$ for GOS. This is to be expected: in fact, the SSC dynamics leave {\em all} the eigenspace of the operator to be averaged invariant, while the SMC leaves invariant only the ones corresponding to maximum and minimal eigenvalue. Preparing only these two while preserving the global expectation allows not only for a better purity, but for a perfect correlation of local measurement outcomes \cite{mazzarella-qconsensus}.
\end{itemize}

\section{Conclusions}
Two classes of QL dynamics that attain asymptotic average consensus for networks of quantum systems have been introduced. The first improves the existing gossip-type dynamics, as in general it attains a purer output state while still guaranteeing SSC consensus. This is achieved by selecting pure representatives for each eigenspace of the global observable whose average has to be preserved. While the new algorithm cannot be simply obtained by randomized permutations of subsystems, as the gossip ones do, it is suitable for a simple discrete-time feedback implementation, in the spirit of \cite{bolognani-arxiv}. The second dynamics are the first proposal of dynamics attaining the stronger $\sigma$SMC consensus, thus not only improve the purity of the output with respect to the gossip-type dynamics, but actually attain a type of consensus that is more similar in spirit to classical consensus: if the local observable $\sigma$ is observed for one subsystem and gives a certain outcome, it is guaranteed that all other subsystems will indeed return the same outcome in subsequent measurements. This second method is also suitable for feedback implementation. Both dynamics ensure convergence in probability when i.i.d. randomization of the application of maps is used, being thus robust with respect to the ordering and suitable for unsupervised implementation. The i.i.d. requirements can be easily relaxed, following e.g. \cite{mazzarella-group}.
Remarkably, the proposed SSC-attaining dynamics can be used, in combination with few additional resources, to asymptotically prepare and stabilize Dicke states in multipartite quantum systems. This can be seen as a discrete-time version of the conditional stabilization protocols proposed in \cite{ticozzi-QLS}  and the results of \cite{ticozzi-ffqls} for continuous time quantum semigroups. Similarly, a strategy can be devised to utilize the SMC-attaining dynamics to prepare GHZ-type states with proper initialization of the network. As a last remark, the presented convergence analysis is one of the first stability results for time-inhomogeneous sequences of CPTP maps using Lyapunov techniques, and could be of inspiration for more general convergence analysis for this important class of dynamics. 

\section*{Acknowledgements}                               
F.T. gratefully acknowledges Lorenza Viola and Peter D. Johnson for joint work on related topics, which made the relevance of Dicke states for consensus apparent, and L. Mazzarella, A. Sarlette and M. Mirrahimi for insightful comments on this work. Work partially supported by the QFuture research grants of the University of Padova, and by the Department of Information Engineering research project QUINTET and QCOS.

\bibliographystyle{plain}
\bibliography{bibQcons3}

\appendix
\section{Discrete-time Invariance Principle}\label{lasalle}
The main tool we are going to use in proving convergence of the SSC algorithm is LaSalle's invariance principle, which we recall here in its discrete-time form \cite{lasalle-discrete}.

\begin{thm}[Discrete-time invariance principle] \label{th:LaSalle}
	Consider a discrete-time system
	$
	x(t+1) = \mathcal{T}[x(t)].
	$
	Suppose $V$ is a $\mathcal{C}^1$ function of $x\in \mathbb R^n$, bounded below and satisfying
	\begin{equation}
	\Delta V(x) = V(\mathcal{T}[x]) - V(x) \le 0, \quad \forall x \label{eq:LaSalleHypotesis}
	\end{equation}
	i.e. $V(x)$ is non-increasing along forward trajectories of the plant dynamics.
	Then any bounded trajectory converges to the largest invariant subset $W$ contained
	in the locus $E=\{x | \Delta V(x) = 0\}$.
\end{thm}

\section{Randomization Lemma}
The following Lemma allows us to conclude the convergence in probability of the proposed methods, as soon as they admit a finite sequence that ensures a strict contraction of the a Lyapunov function.
\begin{lemma}[Convergence in probability]\label{lem:randconv}
Consider a finite number of CPTP maps $\{\Ei_j\}_{j=1}^M,$ and a (Lyapunov) function $V(\rho),$ such that $V(\rho)\geq 0$ and $V(\rho)=0$ if and only if $\rho\in{\cal S},$ with $\cal S$ some density operator set. Assume furthermore:
\begin{enumerate}
\item For each $j$ and state $\rho$, $V(\Ei_j(\rho))\leq V(\rho)$;
\item There exists a finite sequence of maps \beq\label{eq:contrseq}\Ei_o=\Ei_{j_K}\circ \ldots\circ \Ei_{{j_1}},\eeq
with $j_\ell\in\{1,\ldots, m\}$ for all $\ell$, such that $V(\Ei^o(\rho))\leq V(\rho)$ for all $\rho\neq{\cal S}.$
\end{enumerate}
Assume that the maps are selected at random at each time $t$, with independent probability distribution $\Prob_t[\Ei_{j}]$ at each time, such that exist some $\varepsilon>0$ for which $\Prob_t[\Ei_{j}] > \varepsilon$ for all $t$. 
Then, for any $\gamma>0$, the probability of having $V(\rho(t))<\gamma$ converges to $1$ as $t$ converges to $+\infty$.
\end{lemma}

\proof LaSalle-Krasowsii invariance theorem (in discrete time, \cite{lasalle-discrete}) ensures us that by repeating the map $\Ei_o$ infinite time  would entail convergence to $\rho_d,$ and hence $V(\rho)\rightarrow 0.$ Thus, for any fixed $\gamma,$ application of the sequence \eqref{eq:contrseq} for a large enough number of times $N_\gamma > 0$ would ensure $V(\Ei_o^{N_\gamma}(\rho))\leq \gamma$ for any $\rho.$ 

The proof is concluded by noting that the probability that a randomly selected sequence of $B\cdot N_\gamma\cdot K$ maps {\em does not} contain $N_\gamma$ contracting subsequence \eqref{eq:contrseq} is at most $(1-\varepsilon^{N_\gamma\cdot K})^B$. The latter converges to $0$ as $B,$ and thus $t,$ tends to $+\infty$. 
\qed

\end{document}